\documentclass[aps,superscriptaddress,double-spaced,floatfix,showkeys,showpacs]{revtex4}
\usepackage{graphicx,amsmath,bbm,bm,array,subfigure}
\usepackage{appendix}
\usepackage{lipsum}
\usepackage[linktocpage=true,colorlinks=true,linkcolor=blue,citecolor=blue,urlcolor=blue]{hyperref}
\usepackage{MnSymbol}
\def\nn{\nonumber\\}

\begin{document}
\title{Dispersion and suppression of sound near QCD critical point}
\author{Md Hasanujjaman}
\email{jaman.mdh@gmail.com}
\affiliation{Department of Physics, Darjeeling Government College, Darjeeling- 734101, India}
\author{Mahfuzur Rahaman}
\email{mahfuzurrahaman01@gmail.com }
\affiliation{Variable Energy Cyclotron Centre, 1/AF Bidhan Nagar, Kolkata- 700064, India}
\affiliation{Homi Bhabha National Institute, Training School Complex, Mumbai - 400085, India}
 \author{Abhijit Bhattacharyya}
\email{abhattacharyyacu@gmail.com}
\affiliation{Department of Physics,University of Calcutta, 92, A.P.C. Road, Kolkata-700009, India}
\author{Jan-e Alam}
\email{jane@vecc.gov.in}
\affiliation{Variable Energy Cyclotron Centre, 1/AF Bidhan Nagar, Kolkata- 700064, India}
\affiliation{Homi Bhabha National Institute, Training School Complex, Mumbai - 400085, India}

\def\zbf#1{{\bf {#1}}}
\def\bfm#1{\mbox{\boldmath $#1$}}
\def\hf{\frac{1}{2}}
\def\sl{\hspace{-0.15cm}/}
\def\omit#1{_{\!\rlap{$\scriptscriptstyle \backslash$}
{\scriptscriptstyle #1}}}
\def\vec#1{\mathchoice
        {\mbox{\boldmath $#1$}}
        {\mbox{\boldmath $#1$}}
        {\mbox{\boldmath $\scriptstyle #1$}}
        {\mbox{\boldmath $\scriptscriptstyle #1$}}
}
\def \beq{\begin{equation}}
\def \eeq{\end{equation}}
\def \beqa{\begin{eqnarray}}
\def \eeqa{\end{eqnarray}}
\def \nn{\nonumber}
\def \pd{\partial}
\begin{abstract}
We have used second order relativistic hydrodynamics equipped with equation of state which
includes the critical point to study the propagation of perturbation in a relativistic QCD fluid. 
Dispersion relation for the sound wave has been derived to ascertain the fate of the perturbation 
in the fluid  near the QCD critical end point (CEP). We observe that the threshold value of the 
wavelength of the sound in the fluid diverges at the CEP, implying that all the 
modes of the perturbations are dissipated at this point.  Some consequences of the 
suppression of sound near the critical point have been discussed.

\end{abstract}
\maketitle
\section{Introduction}
Relativistic heavy ion collision experiments (RHIC-E) are carried out to create a new state of 
strongly interacting matter, called Quark-Gluon Plasma (QGP)\cite{qgp1,qgp2}, where color degrees of freedom are 
deconfined from their parent hadrons and its properties are governed by the colored 
quarks and gluons. 
The study of the transition from QGP to hadron phase is one of the main goals of RHIC-E. 
For last several years, a lot  of works have been done to explore the QCD phase diagram in 
the $T\,-\mu$ plane where $T$ and $\mu$ denote temperature  and 
baryonic chemical potential respectively. Lattice QCD simulations shows that, at 
vanishing baryon chemical potential ($\mu =0$), the transition from  hadron  to  QGP
is a crossover \cite{fodorandkatz,asakawach,phasediagram,philippe,Aoki} whereas, at 
large $\mu$, 
the transition from hadronic matter to QGP is  found to be first order 
\cite{philippe,Endrodi}. Therefore, it is expected that the first order 
phase transition ends at some point in the $\mu-T$ plane which is called 
the Critical End Point (CEP). The existence of CEP was suggested theoretically in 
Refs.\cite{phasediagram,Berges,Barducci,Kiriyama} and predicted later in lattice simulation 
\cite{Katz,Rajagopal,Fodor}.  The experimental search for the CEP has been
taken up through the beam energy scan (BES) programme at  Relativistic Heavy Ion Collider (RHIC). 
The search will continue in future experiments at 
Facility for Anti-proton and Ion Research (GSI-FAIR) and  Nuclotron-based Ion Collider fAcility (JINR-NICA) rigorously~\cite{cbmbook}.

A major issue in the exploration of the phase digram of QCD is to find out the location of the CEP.
The exact location of the CEP is not known theoretically because
of the difficulties associated with the sign problem of Dirac fermion~\cite{Forcrand,Ding,Gavai} 
in Lattice QCD calculation.  
Some of the QCD based effective models such as NJL, PNJL predict the location of the CEP\cite{PNJL}
with uncertainties ranging from 266-504 MeV in 
$\mu_{c}$ and  115-162 MeV in $T_c$. Therefore, location of CEP in QCD 
phase diagram remains as a big challenging task.  It is one of the  main aim of RHIC-BES programme
\cite{bes1,bes2} to find the CEP by the creating systems 
with different $\mu$ and $T$ by tuning 
the colliding energy, ($\sqrt{s_{NN}}$) of the nuclei. 
At the CEP the correlation length 
diverges~\cite{kunihiro,stephanov1,stephanov2} 
resulting in divergences in several thermodynamic quantities which
may affect signals of QGP. The chances of detecting such effects become greater
if the freeze out curve in $\mu-T$ plane is sufficiently close to the CEP.

In the present work, however, we are not into the search of the location of CEP. 
Rather, we want to examine its effects on the fate of the sound wave propagating through the 
fluid in presence of CEP. Here the  location of the CEP is taken at:
$(T_c, \mu_{c})= (154 \text{MeV}, 367 \text{MeV})$ 
\cite{Asakawa}. It is expected that a system conducive to study the effects of CEP  
may be realised through nuclear collisions at GSI-FAIR,  NICA BES-RHIC. 
The QGP produce in such collisions will expand rapidly along
a trajectory with $s/n$ constant ($s$ and $n$ stand for the entropy density and baryon 
number density respectively) 
and cools down consequently. It is assumed that the isentropic trajectory followed
by  QGP in the $\mu - T$ plane will pass through trajectories which
are very close to the CEP.

The space time evolution of the QGP can be modelled  by the relativistic viscous hydrodynamics.
The first order theory of relativistic viscous hydrodynamics governed by Navier-Stokes (NS) equations  
depends on the first order in dissipative fluxes which is known to violate causality and 
gives unstable solutions \cite{Hiscock}. 
Therefore, making it unsuitable for the description of QGP.
These problems were cured by  Muller \cite{Muller} and Grad \cite{Grad} after including
quantities in second order dissipative flux and
therefore, these theories are called the 'second order hydrodynamics'. 
The relativistic generalization is due to Israel and Stewart \cite{IS} 
which can be used to describe the space-time evolution of   
QGP. The response of the QGP fluid to the perturbation is dictated by
the relevant transport coefficients (shear and bulk viscosities, thermal conductivity, etc.)
of the fluid.  The effects of thermal conductivity $(\kappa)$ and the  
shear viscosity $(\eta)$ have been considered here 
to investigate the propagation of acoustic wave when the system passes 
through the CEP. The effects of CEP in the hydrodynamic evolution enters through the 
Equation of State (EoS).  The EoS is constructed based on the hypothesis
that the transition from QGP to hadrons belongs to the same universality class as 
that of the 3D Ising model. 
The behaviour of thermodynamic quantities 
near CEP is  governed by the critical 
exponents.  Dispersion relation {\it i.e.} the functional dependence of the
frequency ($\omega$) on the  wave vector ($k$) will be set up
to study the effects of CEP on the propagation of the sound wave  in the fluid.

The present work is organized as follows. In section II we will discuss the EoS 
which includes the CEP.  In section III formulation for the propagation of the acoustic wave 
is presented.  The dispersion relation is discussed in section IV. 
Results are presented in section V and section VI is devoted to summary and discussions.
The space like Minkowski metric 
$g^{\mu\nu} =(-,+,+,+)$ and the natural unit  i.e., $c= \hslash=k_B=1$ 
have been used in this work.\\

\section{Equation of State}
The CEP in QGP-hadron transition belongs to the same universality class as that of the  3D Ising model, 
thus a mapping onto QCD phase diagram from the Ising model calculation can be performed. 
It can be shown~\cite{Asakawa,Wathid,Stanley} that the critical entropy density ($s_c$) in QCD is analogous 
to the magnetization ($M$) in  3D Ising model. 
The parameter plane in  3D Ising model are: $r=\frac{T-T_c}{T_c}$ (reduced temperature) and the 
strength of the magnetic field $(\mathcal{H})$.
The CEP in 3D Ising model is located at $(r,\mathcal{H})=(0,0)$. 
Thus $r<0$ represents first order phase transition and $r>0$ signifies crossover transition.  
A critical region is being assumed with linear mapping from the $(r,\mathcal{H})$ to ($\mu,T$) 
plane. The mapping is implemented through the relation:
\begin{eqnarray}
r=\frac{\mu- \mu_{c}}{\Delta \mu_{c}}; \,\,\,\ 
\mathcal{H}=\frac{T- T_c}{\Delta T_c}
\label{eq1}
\end{eqnarray}
where ($T_c, \mu_{c})$ is the location of the CEP
as mentioned above. $\Delta T_{c}$ and $\Delta \mu_{c}$ are chosen as 
elongations of the critical region along T and $\mu $ axis respectively. 
The critical entropy density can be written as 
\begin{eqnarray}
s_c= \frac{M (r,\mathcal{H})}{\Delta T_c}= M\Big (\frac{T- T_c}{\Delta T_c}, \frac{\mu- \mu_{c}}{\Delta \mu_{c}} \Big ) \frac{1}{\Delta T_c}
\label{eq2}
\end{eqnarray}
Firstly, a dimensionless entropy density is constructed as
\begin{eqnarray}
S_c= A (\Delta T_c, \Delta \mu_{c})s_c (T, \mu)
\label{eq3}
\end{eqnarray}
where $A $ is defined as 
\begin{equation}
A(\Delta T_{c},\Delta \mu_{c}) = B \sqrt{\Delta T^2_{c}+\Delta \mu_{c}^2)}
\label{eq4}
\end{equation}
and $B$ is a dimensionless quantity, represents the spread of the critical region. 
In this work we have used  
$(T_c, \mu_c)= (154 \text{MeV}, 367 \text{MeV})$ 
with $(\Delta T_{c}, \Delta \mu_{c},B)=(0.1\,\text{GeV}, 0.2\, \text{GeV}, 2)$.
Using $S_c$ as a switching function, the full entropy density is constructed 
by making a bridge between the entropy density of QGP ($s_Q$) and the hadron ($s_H$) phases. 
The result reads as: 
\begin{eqnarray}
S (T, \mu)=\frac{1}{2}[1-{\rm tanh} \ S_c (T,\mu)] s_{Q}(T,\mu) + \frac{1}{2}[1+{\rm tanh} \ S_c (T,\mu)] s_H(T,\mu)
\label{eq5}
\end{eqnarray}
$s_{Q}$ is calculated by ~\cite{Wathid,satarov}
\beqa
s_{Q}(T,\mu)=\frac{32+21N_{f}}{45}\pi^{2}T^{3}+\frac{N_{f}}{9}\mu^{2}T 
\label{eq6}
\eeqa
where $N_{f}$ is the number of flavour of quarks.\\
$s_H$ can be estimated from the following expression ~\cite{pbraun}, 
\beqa
s_{H}(T,\mu_{B)}=\pm \sum_{i}\frac{g_{i}}{2\pi^{2}}\int^{\infty}_{0}p^{2}\Big[ln\Big(1\pm \{exp(E_{i}-\mu_{i})/T\}\Big)\pm \frac{E_{i}-\mu_{i}}{T\{exp(E_{i}-\mu_{i})/T\pm 1\}}\Big]
\label{eq7}
\eeqa
where the sum is taken over all hadrons with mass up to 2.5 GeV~\cite{Sarwar},
$g_i$ is the statistical degeneracy 
and $E_{i}=\sqrt{p^{2}_{i}+m^{2}_{i}}$ is the energy 
of the $i^{\text{th}}$ hadrons.

Once entropy density is known the thermodynamic quantities such as baryon number density, pressure 
and energy density can be evaluated as follows. The net baryon number density ($n$) is given by:
\begin{eqnarray}
n(T, \mu)= \int_{0}^{T} \frac{\partial S (T^{'}, \mu)}{\partial \mu} dT^{'}
\label{eq8}
\end{eqnarray}
To get the first order phase boundary, we need to take into account the discontinuity 
in the entropy density along the transition line. We add the following term 
to the above equation to  take  this possibility into account (for $T>T_{c}$):
\beqa
\Big|\frac{\pd T_{c}(\mu)}{\pd \mu}\Big|\Big[S(T_{c}+\delta,\mu)-S(T_{c}-\delta,\mu)\Big]
\label{eq9}
\eeqa
where $\Big|\frac{\pd T_{c}}{\pd \mu}\Big|=tan\theta_{c}$ is the tangent 
at the $T_{c}$ and $\delta$ is the small temperature deviation from $T_{c}$.
The pressure can be calculated as:
\begin{eqnarray}
p(T, \mu)=  \int_{0}^{T} S (T^{'}, \mu) dT^{'} 
\label{eq10}
\end{eqnarray}
Finally, the energy density is given by, 
\begin{eqnarray}
\epsilon(T, \mu)= Ts(T,\mu)-p(T,\mu)+\mu n
\label{eq11}
\end{eqnarray}
\section{Propagation of the perturbation in viscous fluid}
IS second order hydrodynamics is appropriate to study
the relativistic fluid nature of QGP as the first order theory (relativistic
NS) violates causality and introduce instability in the solution.
Therefore, in this section we study the propagation of perturbations 
through viscous fluid by using second order causal hydrodynamics. 

One of the major
difference between relativistic and non-relativistic fluid originates  
from the definition of chemical potential. In non-relativistic
case chemical potential constraint the total number of particles in the 
system. But in a relativistic system, the total number of
particles does not remain constant due annihilation and creation
of particles within the fluid. However, through the annihilation and creation 
processes the
conservation of certain quantum numbers remain intact. For example
in strong interaction net (baryon-antibaryon) baryon number, 
net electric charge,  net strangeness
remain conserved (although strangeness is not conserved in weak interaction). 
The present study is concerned with the strong interaction. Accordingly the 
net baryon number will remain conserved throughout the evolution 
of the QGP.  Therefore, in the discussion below the net charge density stands
for net baryon number density.
 
The relativistic energy-momentum tensor ($T^{\lambda\mu}$) in the  
Israel-Stewart  second order hydrodynamics is given by \cite{IS}  
\begin{eqnarray}
T^{\lambda \mu} = \epsilon u^{\lambda}u^{\mu} + P\Delta^{\lambda \mu}+ 2h^{(\lambda}u^{\mu)} +\tau^{\lambda \mu}
\label{eq12}
\end{eqnarray}
where $u^\mu $ is the hydrodynamic four velocity subjected to the normalization condition  
$u^\mu u_\mu =-1,$  $P$ is thermodynamic pressure. The dissipative viscous 
stress $\tau ^{\lambda\mu}=\Pi \Delta^{\lambda\mu}+\pi^{\lambda\mu}$, where $\Pi$ is the bulk viscous
pressure, $\Delta^{\lambda\mu} =g^{\lambda\mu}+u^\lambda u^\mu $  is the spatial projection 
tensor orthogonal to $u^\lambda$ and $\pi^{\lambda\mu }$ is the shear viscous stress with 
$\pi^\lambda_\lambda=h^\lambda u_\lambda=\tau^{\lambda\mu }u_\lambda=0 $.   
The heat flux four vector is  defined as $q^\lambda =h^\lambda-n^\lambda (\epsilon+P)/n$, 
where $n$ is the net baryon number density. The particle four flow is defined as,
\begin{equation}
N^\lambda=nu^\lambda +n^\lambda
\label{eqforN}
\end{equation}
where, $n^{\lambda}$ is called the particle diffusion current, with $n^\lambda u_\lambda=0$. 
The symmetric tensor, $h^{(\lambda}u^{\mu )}$ is defined as $h^{(\lambda}u^{\mu )}=
\frac{1}{2}(h^{ \lambda}u^{\mu}+h^{\mu }u^{\lambda})$. 

The definition of fluid four velocity in Eq.\eqref{eq12} can be fixed 
by choosing a suitable reference frame attached to the fluid element according to Landau-Lifshitz 
(LL)\cite{Landau}  or Eckart\cite{Eckart}. The Eckart frame represents a Local Rest Frame (LRF) for 
which the net charge dissipation is zero but the net energy dissipation is non zero and the LL 
frame represents a local rest frame where 
the energy dissipation is zero but the net charge dissipation is non-zero. 
We consider LL frame here 
to study a system having  non-zero net baryon number density.\\ 
 
In LL frame: $h^\mu=0$, $n^\mu =-n q^\mu /(\epsilon+P)$ 
and the different viscous fluxes are given by \cite{IS},
\begin{eqnarray}
\Pi &=&-\frac{1}{3}\zeta(\pd_{\mu}u^\mu +\beta_0 D \Pi-\alpha_0  \pd_{\mu}q^\mu )\nonumber\\   
\pi^{\lambda \mu}&=&-2\eta \Delta^{\lambda\mu\alpha\beta}\Big[\partial_{\alpha}u_{\beta}+\beta_{2}D\pi_{\alpha\beta}-\alpha_{1}\partial_{\alpha}q_{\beta}\Big]\nonumber\\
q^{\lambda}&=&\kappa T\Delta^{\lambda\mu} [\frac{nT}{\epsilon+P}(\partial_\mu \alpha )-\beta_1 D{q_\mu}+\alpha_0\partial_\mu \Pi +\alpha_1\pd_{\nu}\pi ^{\nu}_{\mu} ] 
\label{eq13}
\end{eqnarray}
where  $D\equiv u^\mu\partial_\mu$, is known as co-moving derivative and in LRF, $D\Pi =\dot{\Pi }$ represents the time derivative.
The double symmetric traceless projection operator is defined by 
$\Delta^{\mu\nu\alpha\beta}=
\frac{1}{2}\big[\Delta^{\mu\alpha}\Delta^{\nu\beta}+\Delta^{\mu\beta}\Delta^{\nu\alpha}-
\frac{2}{3}\Delta^{\mu\nu}\Delta^{\alpha\beta}\big]$, 
$\Delta^{\mu\nu}\pd_{\nu}=\nabla^{\mu}$ and $\Delta^{\mu\nu}u_{\mu}=0$.  
The quantity $\alpha=\mu/T$ appearing in Eq.\eqref{eq13} is known as thermal potential and
$\eta$, $\zeta $, $\kappa$ are the coefficients of shear viscosity, bulk viscosity and thermal conductivity respectively, 
$\beta _0,\beta_1,\beta_2$ are relaxation coefficients, 
$\alpha_0$ and $\alpha_1$ are coupling coefficients. 
The relaxation times for the 
bulk pressure ($\tau_{\Pi}$), the heat flux ($\tau_q$) and the shear tensor ($\tau_{\pi}$) 
are defined as~\cite{muronga} 
\begin{equation}
\tau_{\Pi}=\zeta \beta_0,\,\,\,\, \tau_q=T\beta_1,\,\,\,\, \tau _{\pi}=2\eta \beta_2
\label{eq14}
\end{equation}
The relaxation lengths which  couple to   heat flux and bulk  pressure 
($l_{\Pi q}, l_{q\Pi}$), the  heat flux and shear tensor $(l_{q\pi},l_{\pi q})$ 
are defined as, 
\begin{equation}
l_{\Pi q}=\zeta \alpha_0,\,\,\,\, l_{q\Pi}=k_B T\alpha_0,\,\,\,\,
l_{q\pi}=k_BT\alpha_1,\,\,\,\,l_{\pi q}=2\eta \alpha_1 
\label{eq15} 
\end{equation}  
At the ultra-relativistic limit, $\beta=m/T\rightarrow 0$ where $m$ is the mass of the
particle and we have the following relations ~\cite{IS},
\begin{eqnarray}
\alpha_0 \approx  6\beta^{-2}P^{-1}, \alpha_1 \approx -\frac{1}{4}P^{-1},
\beta_0 \approx 216 \beta^{-4} P^{-1},\beta_1 \approx \frac{5}{4}P^{-1}, 
\beta_2 \approx \frac{3}{4}P^{-1} 
\label{eq16}
\end{eqnarray}

Since in energy frame,  $h^\mu=0$, then the energy-momentum tensor(EMT) reduces to 
\begin{equation}
T^{\lambda\mu}=\epsilon u^\lambda u^\mu+P\Delta^{\lambda\mu} +\Pi\Delta^{\lambda\mu}+\pi ^{\lambda\mu} 
\label{eq17}
\end{equation}
Putting the explicit forms of $\Pi, q^\lambda$ and $\pi ^{\lambda\mu}$ given by Eq.\eqref{eq13} into Eq.\eqref{eq12} and keeping only the terms up to second order in space time derivatives, the EMT becomes  \cite{Mahfuzur}
\begin{eqnarray}
 T^{\lambda\mu}&= & \epsilon u^\lambda u^\mu+P\Delta^{\lambda \mu}- \frac{1}{3}\zeta \Delta^{\lambda\mu}\partial_\alpha u  ^\alpha + \frac{1}{9} \zeta \beta_0\Delta^{\lambda\mu} D(\zeta\partial_\alpha u  ^\alpha)   +\frac{\zeta \alpha _0  }{3}\Delta^{\lambda\mu}\partial_\alpha\Big \{\frac{n\kappa T^2}{\epsilon+P}\nabla^\alpha(\alpha) \Big\}\nonumber \\&-&2\eta \Delta^{\lambda\mu\alpha\beta} \partial_\alpha u_\beta  
  + 4\eta \beta_2\Delta^{\lambda\mu\alpha\beta} D(\eta\Delta_{\alpha\beta}^{\rho\sigma} \partial_\rho u_\sigma)  +2\alpha _1\eta\Delta^{\lambda\mu\alpha\beta}\partial_{\alpha}\Big\{\frac{ n\kappa T^2}{\epsilon+P}\nabla_\beta(\alpha)\Big\}
\label{eq18}
\end{eqnarray}

The solution of  IS hydrodynamical equations grants stability and
causality.  This is achieved by promoting the dissipative
currents as independent  dynamical variables and introducing relaxation time scales
for these currents.  In NS theory the 
dissipative currents instantaneously respond to the hydrodynamical gradients but 
in IS theory the response of the dissipative currents is governed by the
relaxation time scales (see Eq.~\eqref{eq14}). The energy-momentum tensor given in
Eq.~\eqref{eq18} represents second order dissipative hydrodynamics which
is equivalent to IS theory for small gradients.
The general form of the EMT constrained by the conformal invariance can be found in ~\cite{Baier}.

The full charge current (up to second-order in velocity gradient) can be written as, 
\begin{eqnarray}
N^\mu &=&nu^\mu -\frac{n\kappa T}{(\epsilon+P)}\Big[ \frac{nT}{(\epsilon+P)} 
\nabla^\mu \alpha  -\beta_1\Delta^{\mu\nu }D\ \Big\{ \frac{n\kappa T^2}{(\epsilon+P)}
\nabla_\nu  \alpha \Big\}-\frac{\alpha_0}{3}
\nabla^\mu(\zeta\partial _\alpha u^\alpha)\nonumber\\
& -&2 \alpha_1\Delta^{\mu\nu } \partial^\rho(\eta\Delta _{\rho\nu }^{\alpha\beta} 
\partial_\alpha u_\beta) \Big]
\label{eqcurrent}
\end{eqnarray}
Eqs.\eqref {eq18} and \eqref{eqcurrent} governs the motion of perturbations in the
relativistic viscous fluid with one conserved current (baryonic current for
the present case). 

We impart small perturbations $P_1,\epsilon_1, n_1, T_1, \mu_1$ and $ u^\alpha_1$  to
$P, \epsilon, n, T,\mu$ and $u^\alpha $  respectively to study 
the propagation of acoustic wave in the fluid with $u^\alpha=(1,0,0,0)$  as outlined in 
Ref.\cite{Weinberg1971}.  
We   set $u^0_1=0$ to preserve the normalization condition $u^\alpha u_\alpha =-1$. 
 
A  space time dependent perturbation $\sim exp[-i(k  x -\omega t)]$  is imparted
to the fluid and its fate is being studied.
The equation of motions that dictate the evolution of different components 
of the perturbations  can be obtained from the 
the conservation  of the energy-momentum tensor ($T^{\mu\lambda}$)  and net-baryon number 
($N^\mu$) of the fluid:
\begin{equation}
\partial_\mu T^{\mu\lambda }=0, \,\,\,\,\, \partial_\mu {N}^\mu=0
\label{eq19}
\end{equation} 
The equation of motion of various components of energy momentum tensor 
are given by: 
\begin{eqnarray}
0&=&\omega T_1^{i0}-k_jT_1^{ij}\nonumber\\
&=&\omega (\epsilon+P)u_1^i-k^iP_1+\frac{1}{3}\zeta k^i\Big[ i(\vec{k}\cdot \vec{u_1})+\frac{1}{3}\zeta\beta _0\omega(\vec{k}\cdot \vec{u_1})\Big] +i\eta\Big[ k^2u^i_1+\frac{1}{3}k^i (\vec{k}\cdot \vec{u_1})\Big]\nonumber\\
&&-2\eta^2 \beta_2 \omega\Big[ k^2u^i_1+\frac{1}{3}k^i (\vec{k}\cdot \vec{u_1})\Big]+\frac{nT\kappa}{(\epsilon+P)}(\mu _1-\alpha T_1)\Big[ \frac{\alpha _0\zeta k^2}{3}k^i+\frac{4}{3}\alpha_1\eta k^2 k^i\Big]
\label{eq20}
\end{eqnarray}
and the other components of the EMT  satisfies,
\begin{eqnarray}
0&=&\omega T_1^{00}-k_iT_1^{i0}\nonumber
\\&=&\omega\epsilon_1-(\epsilon+P)(\vec{k}\cdot \vec{u_1})
\label{eq21}
\end{eqnarray}
The number conservation equation gives,
\begin{eqnarray}
 0=\omega n_1-n(\vec{k}\cdot\vec{u_1})-i \frac{n^2\kappa T k^2}{(\epsilon+P)^2}\Big(\mu_1-\alpha T_1\Big)\Big(1+i\omega\kappa \beta_1\Big)+\frac{1}{3}\frac{n \kappa Tk^2}{(\epsilon+P)}\Big(\zeta \alpha_0+4 \eta \alpha_1\Big)(\vec{k}\cdot\vec{u_1})
\label{eq22}
\end{eqnarray}
In Eqs.~\eqref{eq20},\eqref{eq21} and \eqref{eq22}, we considered terms upto first 
order in perturbations and neglected the higher order terms. Also, we have not perturbed the 
different transport coefficients, as they are not hydrodynamical variables.
In LRF, we take them as constant in space and time,
hence their comoving derivative are zero. 
For simplicity of calculation we only considered shear viscosity ($\eta$) and thermal conductivity ($\kappa$) and neglected the bulk viscosity ($\zeta$).
We decompose the fluid velocity into directions perpendicular and parallel to
the direction of wave vector, $\vec{k}$ as:\begin{equation}
\vec{u_1}=\vec{u_1}_\bot +\vec{k} (\vec{k}\cdot\vec{u_1})/k^2
\label{eq23}
\end{equation}
The modes propagating along the direction of $\vec{k}$ are called longitudinal 
and those perpendicular to $\vec{k}$ are called transverse modes. 

The quantities, $\epsilon_1$, $P_1$ and $\mu_1$ defined above can be 
expressed in terms of thermodynamic quantities as follows:
\begin{eqnarray}
&&\epsilon_1=\Big(\frac{\partial \epsilon}{\partial T}\Big)_n T_1+\Big( \frac{\partial \epsilon}{\partial n}\Big)_T n_1
\nn\\&&P_1=\Big( \frac{\partial P}{\partial T}\Big)_n T_1+\Big( \frac{\partial P}{\partial n}\Big)_T n_1
\nn\\&&\mu_1=\Big[-\Big( \frac{\partial n}{\partial T}\Big)_n T_1+n_{1}\Big]\Big( \frac{\partial \mu}{\partial n}\Big)_T 
\label{eq24}
\end{eqnarray}

\section{Dispersion Relations}
Eqs.  \eqref {eq20}, \eqref{eq21} and \eqref{eq22} can be used to write
down the algebraic equation satisfied by $\omega$ as, 
\begin{eqnarray}
a\omega^3+b\omega^2+c\omega=0\hspace{0.5cm}\rightarrow\hspace{0.5cm}\omega(a\omega^2+b\omega+c)&=&0 
\label{eq25}
\end{eqnarray}
The coefficients $a, b$ and $c$ are determined by solving 
Eqs. \eqref {eq20}, \eqref{eq21} and \eqref{eq22} simultaneously. 
The solutions of this equation which provide a relation between $\omega$ and $k$
is called the dispersion relation.
The equation, \eqref{eq25} has  
three roots, one real which is 
$\omega=0$ and two complex roots with real ($\omega_{\Re e}$) and imaginary ($\omega_{\Im m}$) 
parts given below by Eqs.~\eqref{eq26} and \eqref{eq27} respectively.
The real part of $\omega$ can be expressed as:
\begin{equation}
\omega_{\Re e}=\sqrt{\frac{a_0k^2-a_1k^3+a_2k^4}{b_0-b_1k^2}}
\label{eq26}
\end{equation}
where
\begin{eqnarray}
&&a_0=9h\left[\left(\frac{\partial P}{\partial T}\right)_n
+\alpha_1n
\left\lbrace\left(\frac{\partial\epsilon}{\partial n}\right)_T
\left(\frac{\partial P}{\partial T}\right)_n
-\left(\frac{\partial\epsilon}{\partial T}\right)_n
\left(\frac{\partial P}{\partial n}\right)_T
\right\rbrace\right]\nn\\
&&a_1=\frac{9\alpha\beta_1n^2T^2\kappa^2}{h}+
12\alpha_1\eta\kappa nT\left[\alpha+\frac{\alpha n}{h}\left(\frac{\partial\epsilon}{\partial n}\right)_T
-\frac{T}{h}\left(\frac{\partial P}{\partial T}\right)_n
+\frac{T}{h}\left(\frac{\partial\epsilon}{\partial T}\right)_n\right]
\nn\\
&&a_2=\frac{9\beta_1\kappa^2n^2T}{h}\left[
\left(\frac{\partial n}{\partial T}\right)_\mu\left(\frac{\partial\mu }{\partial n}\right)_T
+\left(\frac{\partial P}{\partial T}\right)_n\left(\frac{\partial P}{\partial n}\right)_T\right]\nn\\
&&+\frac{12\alpha_1\eta\kappa nT}{h}\left[\left(\frac{\partial P}{\partial T}\right)_n 
+n\left(\frac{\partial n}{\partial T}\right)_\mu\left(\frac{\partial\mu}{\partial n}\right)_T
+\frac{n}{h}\left(\frac{\partial P}{\partial T}\right)_n\left(\frac{\partial\epsilon}{\partial n}\right)_T\right]\nn\\
&&b_0=9h\left(\frac{\partial\epsilon}{\partial T}\right)_n\nn\\
&&b_1=24\beta_2\eta^2\left(\frac{\partial\epsilon}{\partial T}\right)_n+\frac{9\beta_1\kappa^2 n^2}{h}\left[
T\left(\frac{\partial\mu}{\partial n}\right)_T\left(\frac{\partial n}{\partial T}\right)_\mu
-T^2\left(\frac{\partial\epsilon}{\partial n}\right)_T\left(\frac{\partial\mu}{\partial n}\right)_n
+\alpha\left(\frac{\partial\epsilon}{\partial n}\right)_T\right]
\label{eq026}
\end{eqnarray}
and $h=\epsilon+P$ is the enthalpy density.   
We have kept terms up to quadratic power of
transport coefficients in Eq.~\eqref{eq026}.  
We have also neglected the higher order terms in $\alpha_{0},\alpha_{1},\beta_{0},\beta_{1},\beta_{2}$.
Expanding $\omega_{\Re e}$ in powers of $k$ and keeping terms up to $\cal{O}$$(k^4)$ we obtain,
\begin{equation}
\omega_{\Re e}=\sqrt{\frac{a_0}{b_0}}\left[k-\frac{1}{2}\frac{a_1}{a_0}k^2+(\frac{1}{2}\frac{a_2}{a_0}-\frac{1}{8}{a_1}{a_0^2}+\frac{b_1}{b_0})k^3
+(\frac{1}{4}\frac{a_1a_2}{a_0^2}+\frac{1}{16}{a_1^2}{a_0^3}-\frac{1}{2}\frac{a_1b_1}{a_0b_0})k^4\right]
\label{omegaRepower}
\end{equation}

Similarly, the expression for the imaginary part of $\omega$ reads as:
\begin{equation}
\omega_{\Im m}=\frac{-c_0k^2+c_1k^3+c_2k^4}{d_0+d_1k^2}
\label{eq27}
\end{equation}
where
\begin{eqnarray}
&& c_0=2\eta h^2\left(\frac{\partial\epsilon}{\partial T}\right)_n
-3hn^2\kappa T\beta_1\left[\alpha\kappa\left(\frac{\partial\epsilon}{\partial n}\right)_T
+h\left(\frac{\partial n}{\partial T}\right)_\mu
\left(\frac{\partial\mu }{\partial n}\right)_T+\frac{\alpha_1}{\beta_1}\left(\frac{\partial P}{\partial T}\right)_n\left(\frac{\partial\epsilon}{\partial n}\right)_T\right]\nn\\
&&c_1=2\alpha\beta_1\eta n^2 T\left(T\kappa^2+4 h\eta\frac{\beta-2}{\beta_1}\right)\nn\\
\label{eq06}
&&c_2=8\beta_2\eta\kappa n^2T\left[\left(\frac{\partial\epsilon}{\partial T}\right)_n-\left(\frac{\partial P}{\partial n}\right)_{T}^2
\left(\frac{\partial\epsilon}{\partial n}\right)_T\right]\nn\\
&&d_0=3h^3\left(\frac{\partial \epsilon}{\partial T}\right)_n\nn\\
&&d_1=3h\beta_1n^2\kappa\left[\alpha\kappa\left(\frac{\partial\epsilon}{\partial n}\right)_T+4T^2\kappa\left(\frac{\partial\epsilon}
{\partial T}\right)_n\left(\frac{\partial\mu}{\partial n}\right)_T^2-
T\left(\frac{\partial n}{\partial T}\right)_n\left(\frac{\partial\epsilon}{\partial n}\right)_T\right]
\label{eq027}
\end{eqnarray}
The imaginary part of $\omega$ up to $\cal{O}$$(k^4)$  is given by,
\begin{equation}
\omega_{\Im m}=-\frac{c_0}{d_0}\left[k^2-\frac{c_1}{c_0}k^3-(\frac{d_1}{d_0}+\frac{c_2}{c_0})k^4\right]
\label{omegaImpower}
\end{equation}
The dispersion relation for first order hydrodynamics can be obtained 
by setting the relaxation coefficients ($\beta _0,\beta_1,\beta_2$)
and  the coupling coefficients ($\alpha_0$, $\alpha_1$) to zero which allows
only $a_0$, $b_0$, $c_0$ and $d_0$ to be non-zero. Therefore, keeping terms
up to $\cal{O}$$(k^2)$ in Eqs.~\eqref{omegaRepower} and \eqref{omegaImpower} we get
(see also~\cite{Grozdanov}),
\begin{equation}
\omega (k)=c_s k -\frac{i}{2}k^{2}\frac{\eta}{s}s\frac{4/3}{h}
\label{eq28}
\end{equation}
where $c_{s}=\sqrt{\big(\frac{\pd p}{\pd \epsilon}\big)_{s/n}}$ is the speed of sound and $\eta/s$ is 
shear viscosity to entropy density ($s$) ratio. The Eq.\eqref{eq28} is the dispersion relation for
NS hydrodynamics.

\subsection{Fluidity near the critical region}
The imaginary and real parts of $\omega$  provide the information 
respectively on attenuation and the propagation of the sound wave in the dissipative fluid.
Thus, if   magnitude of the imaginary part  is larger than the real part, 
the wave will dissipate quickly. 
The dispersion relation in Eqs.\eqref{eq26} and \eqref{eq27} 
can be used to determine the upper limit of $k$ of the sound wave that 
will dissipate in the medium. The threshold value of $k$, $k_{th}$  can be calculated by using the 
following condition~\cite{liao}
\begin{eqnarray}
\Big|\frac{\omega_{\Im m}(k)}{\omega_{\Re e}(k)} \Big|_{k=k_{th}}=1
\label{eq29}
\end{eqnarray}
{\it i.e.} any wave with wave vector higher than $k_{th}$ will get
dissipated in the fluid.
Solving the above equation, we get, 
\begin{equation}
k_{th}=\sqrt{\frac{\mathcal{P}}{\mathcal{Q}}}\nn\\
\label{kth}
\end{equation}
where
\begin{eqnarray}
\mathcal{P}&=&\frac{a_{0}}{c_{0}^{2}}-\frac{a_{0}d_{1}c^{2}_{2}}{b_{0}c_{0}^{2}d_{0}^{2}}-
\frac{a_{1}d_{0}}{c_{0}^{3}}+\frac{a_{1}c_{2}^{2}d_{1}}{b_{0}c_{0}^{3}d_{0}}-
\frac{b_{1}^{2}c_{2}^{3}d_{1}}{b_{0}c_{0}^{2}d_{0}^{3}}+\frac{a_{1}d_{0}}{a_{0}b_{0}c_{0}^{2}}
\end{eqnarray}
and
\begin{eqnarray}
\mathcal{Q}&=&\frac{b_{0}}{d_{0}^{2}}-\frac{b_{1}}{c_{0}^{2}}-
\frac{b_{0}c_{2}}{c_{0}^{3}}+\frac{a_{1}}{a_{0}b_{0}d_{0}}-
\frac{a_{1}b_{1}d_{0}}{a_{0}b_{0}^{2}c_{0}^{2}}+
\frac{a_{1}c_{2}d_{0}}{b_{0}^{2}c_{0}^{3}}
\end{eqnarray}
Expanding $\mathcal{P}$ and $\mathcal{Q}$ and keeping the first term of the series we get, 
\begin{eqnarray}
k_{th}&=&\sqrt{\frac{a_{0}d_{0}^{2}}{b_{0}c_{0}^{2}}}\Bigg[1-\frac{1}{2}\Bigg(\frac{d_{1}c_{2}^{2}}{b_{0}d_{0}^{2}}+
\frac{a_{1}d_{0}}{a_{0}c_{0}}-\frac{a_{1}c_{2}^{2}d_{1}}{a_{0}b_{0}c_{0}d_{0}}+\frac{b_{1}^{2}c_{2}^{3}d_{1}}{a_{0}b_{0}d_{0}^{3}}
-\frac{a_{1}d_{0}}{a_{0}^{2}b_{0}}\Bigg)\Bigg]\nonumber\\
&&\Bigg[1+\frac{1}{2}\Bigg(\frac{b_{1}d_{0}^{2}}{b_{0}c_{0}^{2}}+\frac{c_{2}d_{0}^{2}}{c_{0}^{3}} -\frac{a_{1}d_{0}}{a_{0}b_{0}^{2}}+\frac{a_{1}b_{1}d_{0}^{3}}{a_{0}b_{0}^{3}c_{0}^{2}}-\frac{a_{1}c_{2}d_{0}^{3}}{b_{0}^{3}c_{0}^{3}}\Bigg)\Bigg]
\label{eq31}
\end{eqnarray}
The first term of the above expression gives the value of $k_{th}$ in the NS limit, 
$k_{th}=\sqrt{\frac{a_{0}d_{0}^{2}}{b_{0}c_{0}^{2}}}=\frac{3}{2}c_s\frac{h}{s}\frac{1}{\eta/s}$. 
The subsequent terms arise from the second order hydrodynamical effects as indicated 
by the presence of  coupling and relaxation coefficients appearing through the
quantities defined in Eqs.~\eqref{eq026} and \eqref{eq027}.

The wavelength ($\lambda_{th}$) corresponding to $k_{th}$
is given by  
$\lambda_{th}= 2\pi/k_{th}$. Sound waves with  wavelength, 
$\lambda<\lambda_{th}$ will dissipate in the medium.
However, sound wave with $\lambda>\lambda_{th}$ will propagate in the fluid
without much dissipative effects. The quantity $\lambda_{th}$ can be used to define the 
fluidity of fluids with widely varying particle density and temperature
by selecting a length scale (inter-particle separation), $l\sim \rho^{-1/3}$~\cite{liao} of the 
system as: 
\begin{equation}
\mathcal{F}\sim \frac{\lambda_{th}}{l}.
\label{eq32}
\end{equation}
where $\rho$ is the particle number density of the fluid (for relativistic
fluid $l$ can be chosen as $l\sim s^{-1/3}$). 
The length scale $R_v\sim 1/k_{th}$, called viscous
horizon \cite{staig} sets the limit for sound 
with $\lambda$ smaller than $R_v$ will be 
dissipated due to viscous 
and thermal conduction effects. $R_v$ can be used to estimate
the value of the highest 
harmonics $n_v=2\pi R/R_v$  which will 
survive the dissipation {\it i.e.} any harmonics of order higher
than $n_v$ will not survive against dissipation. 
We find that $k_{th}$ is directly proportional to the speed of sound ($c_{s}$), which approaches 
zero near critical point. Therefore, we can argue that $k_{th}$ also vanishes 
or in other words $\lambda_{th}$ diverges at the critical point.
\section{Results and Discussions} 
\begin{figure}[h]
\centering
\includegraphics[width=8.5cm]{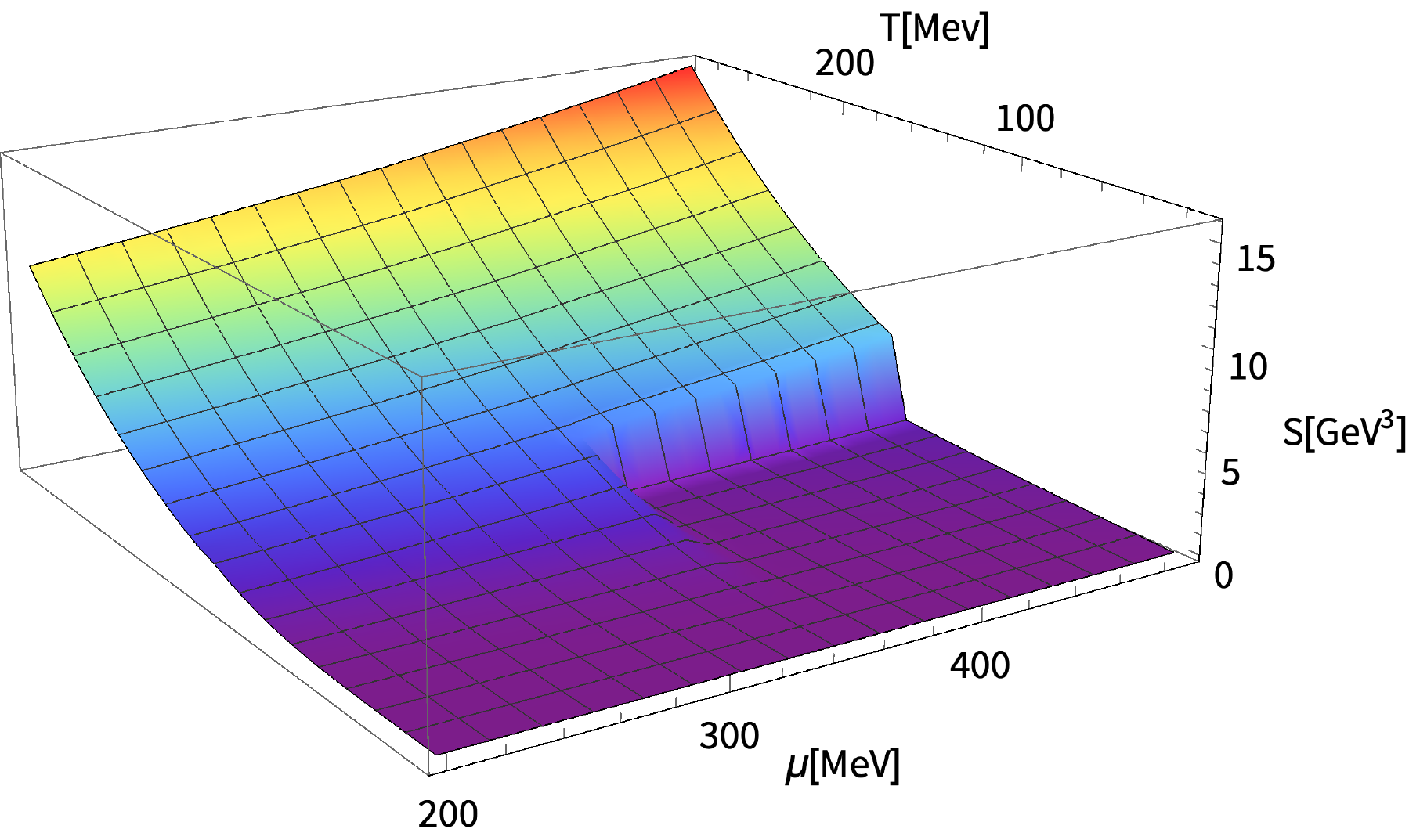}
\includegraphics[width=8.5cm]{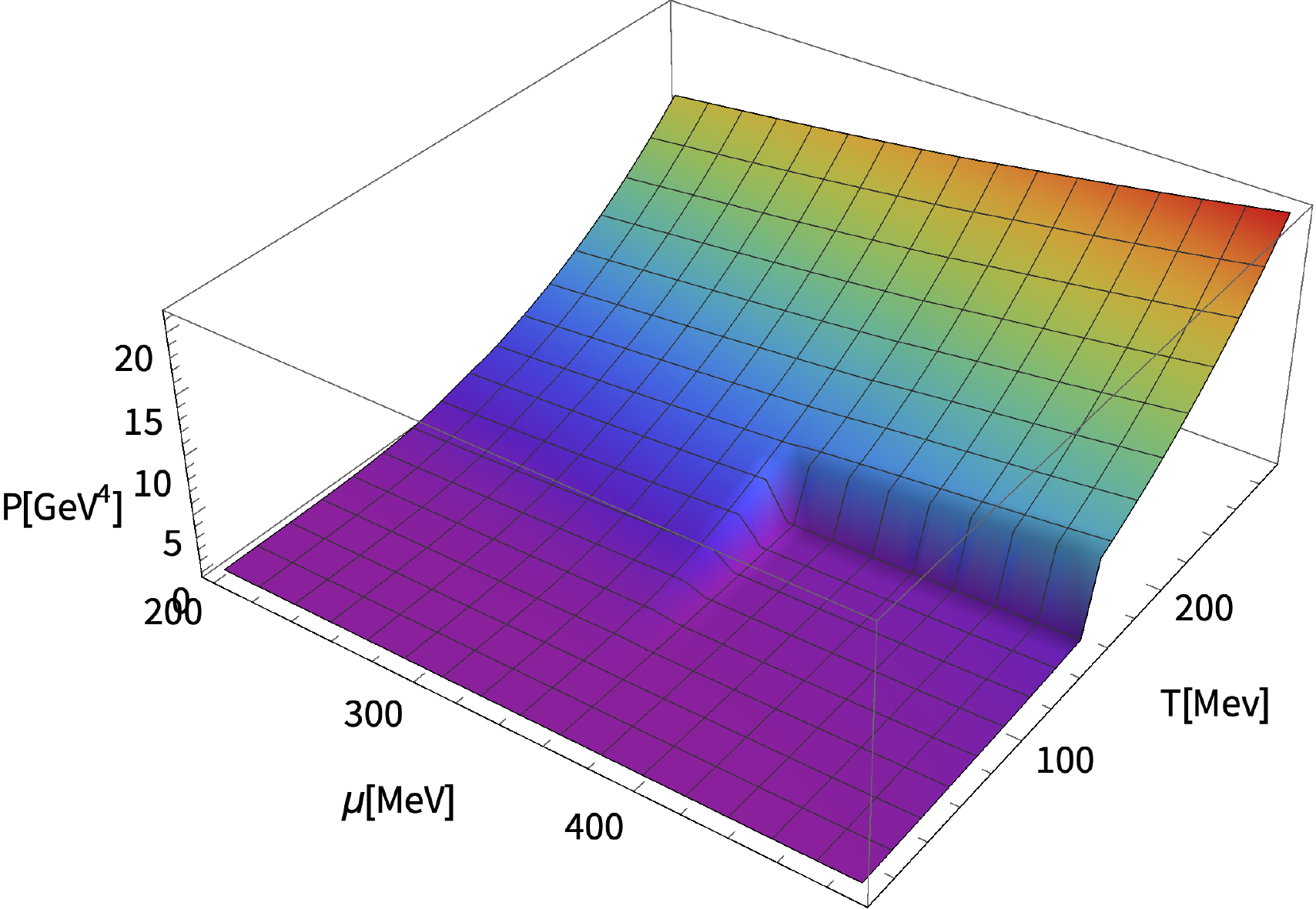}		
\caption{(color online) Left panel is the constructed entropy density as a function of ($T$, $\mu$). Right 
panel is the pressure as a function of ($T$, $\mu$). We consider CEP is at $(T_c, \mu_{c})= (154 ,367) MeV$ .}
\label{fig1}
\end{figure}

In Fig.~\ref{fig1} the variation of entropy density (left panel) and pressure (right panel)
with $\mu$ and $T$ have been depicted. The EoS includes the critical point
at $(T_{c},\mu_{c})$=(154 MeV, 367 MeV).  
The effects of critical point is clearly visible on the entropy density and pressure. 
The discontinuity in entropy density at large baryonic
chemical potential ($\mu$) is indicating a first order phase transition.
(left panel, Fig.~\ref{fig1}). 

\begin{figure}[h]
\centering
\includegraphics[width=8.5cm]{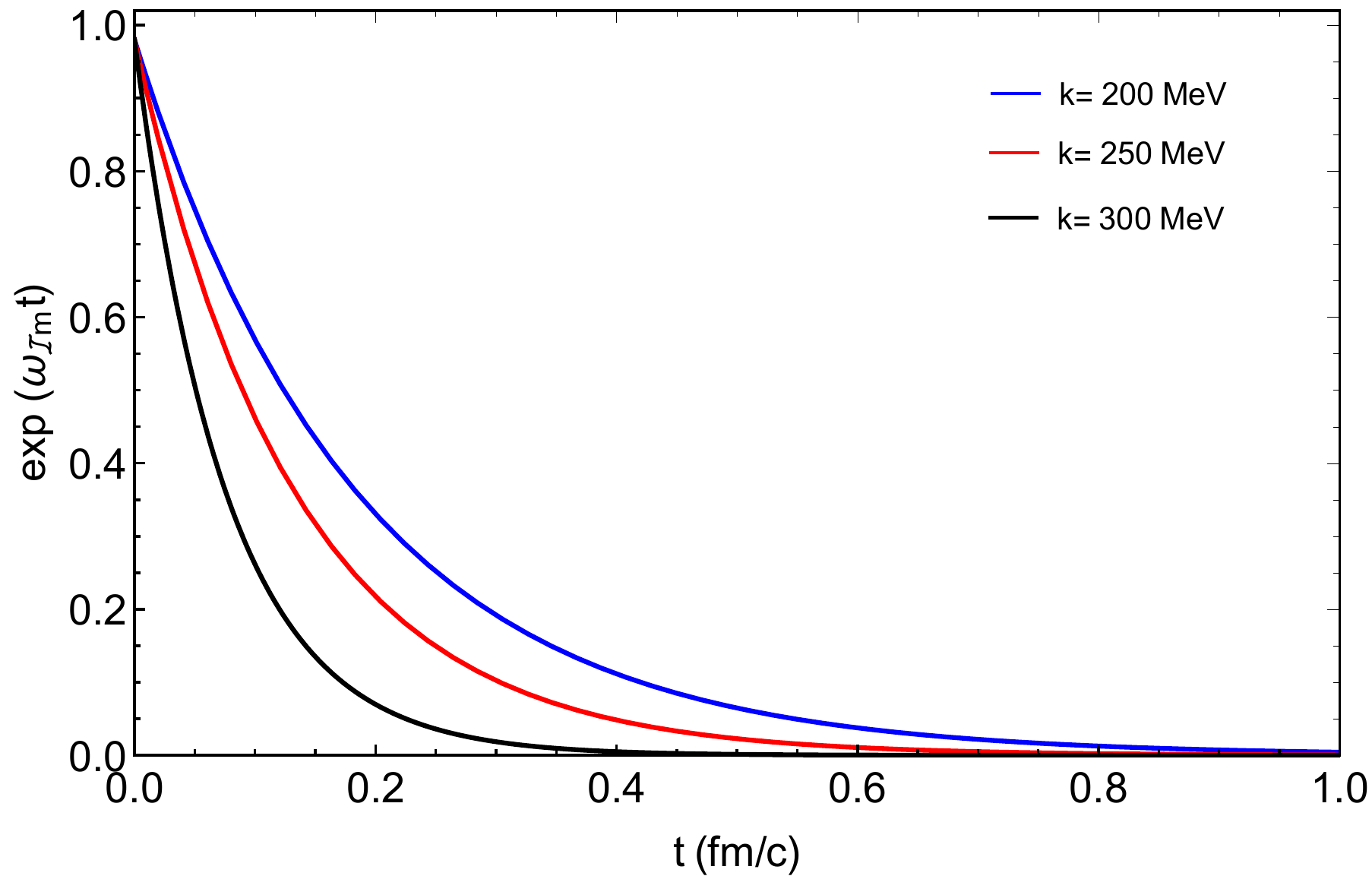}
\includegraphics[width=8.5cm]{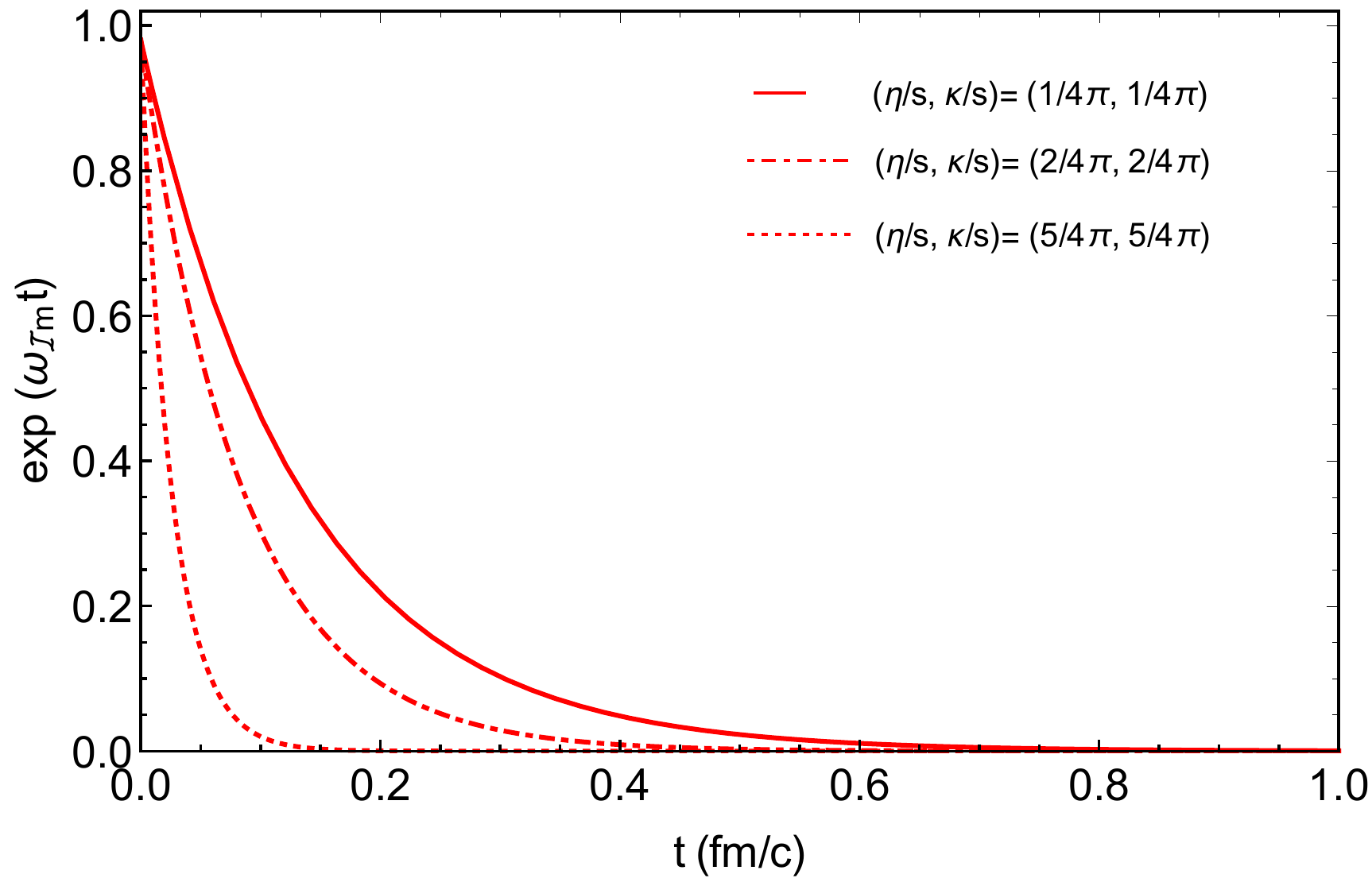} 
\caption{(color online) a) Dissipation of sound modes with time for different k values  at $(T_{c},\mu_{c})=(154, 367)$
MeV and ($\eta/s=\kappa/s=1/4\pi).$  b) Damping of the sound waves with different sets of transport 
coefficients with $k= 250 MeV$.}
\label{fig2}
\end{figure}
\begin{figure}[h]
	\centering
		\includegraphics[width=8.5cm]{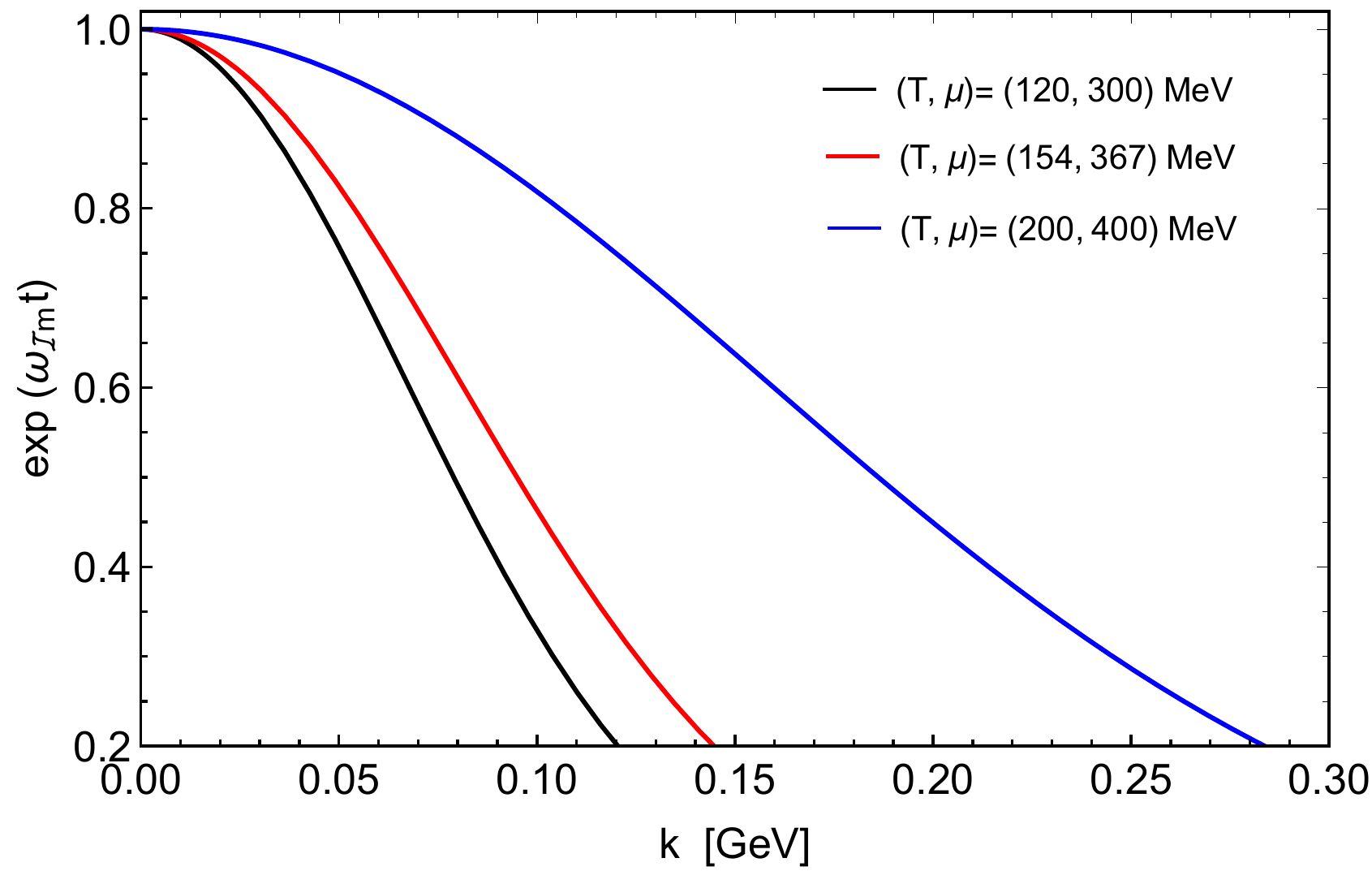}
		\includegraphics[width=8.5cm]{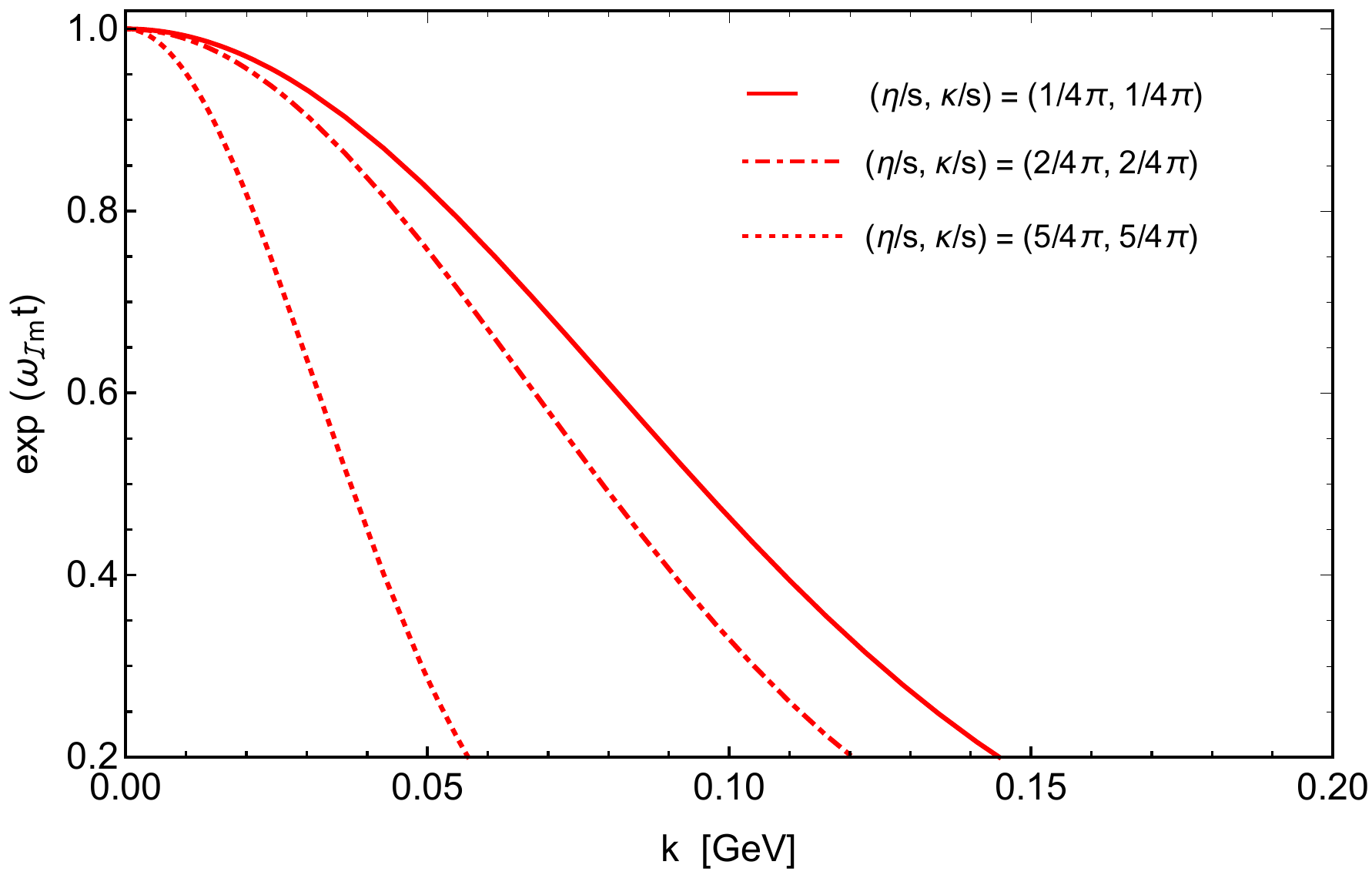} 
		\caption{(color online) a) Dissipation of sound waves with k for different sets of $(T,\mu)$ at $t = 0.6 fm/c$ and ($\eta/s=\kappa/s=1/4\pi)$. b) Damping of the sound waves with different sets of value of transport coefficients with $(T_{c},\mu_{c}) = (154,367)$ MeV.}
\label{fig3}
\end{figure}
\begin{figure}[h]
	\centering
		\includegraphics[width=8.5cm]{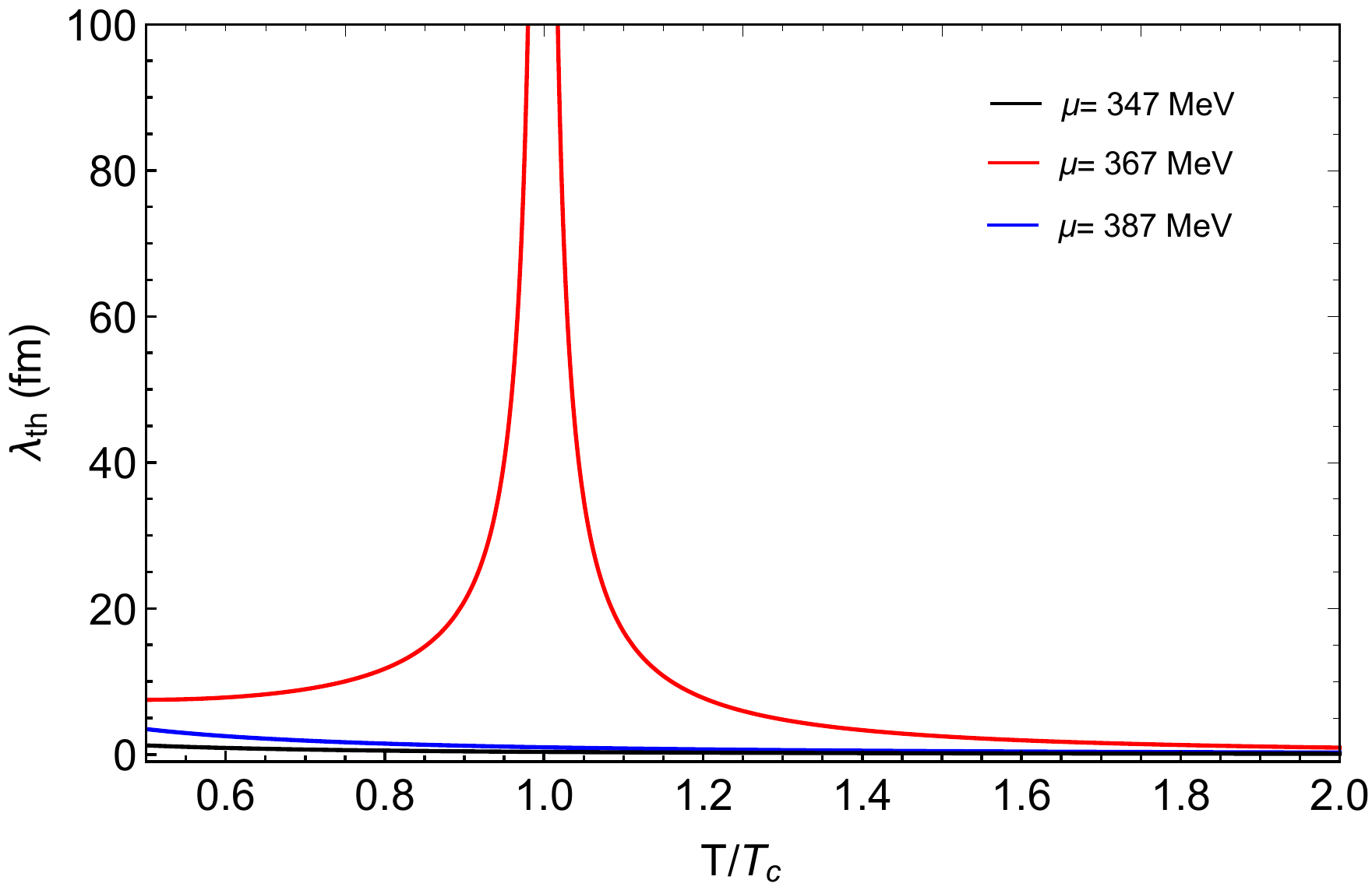}
			\includegraphics[width=8.5cm]{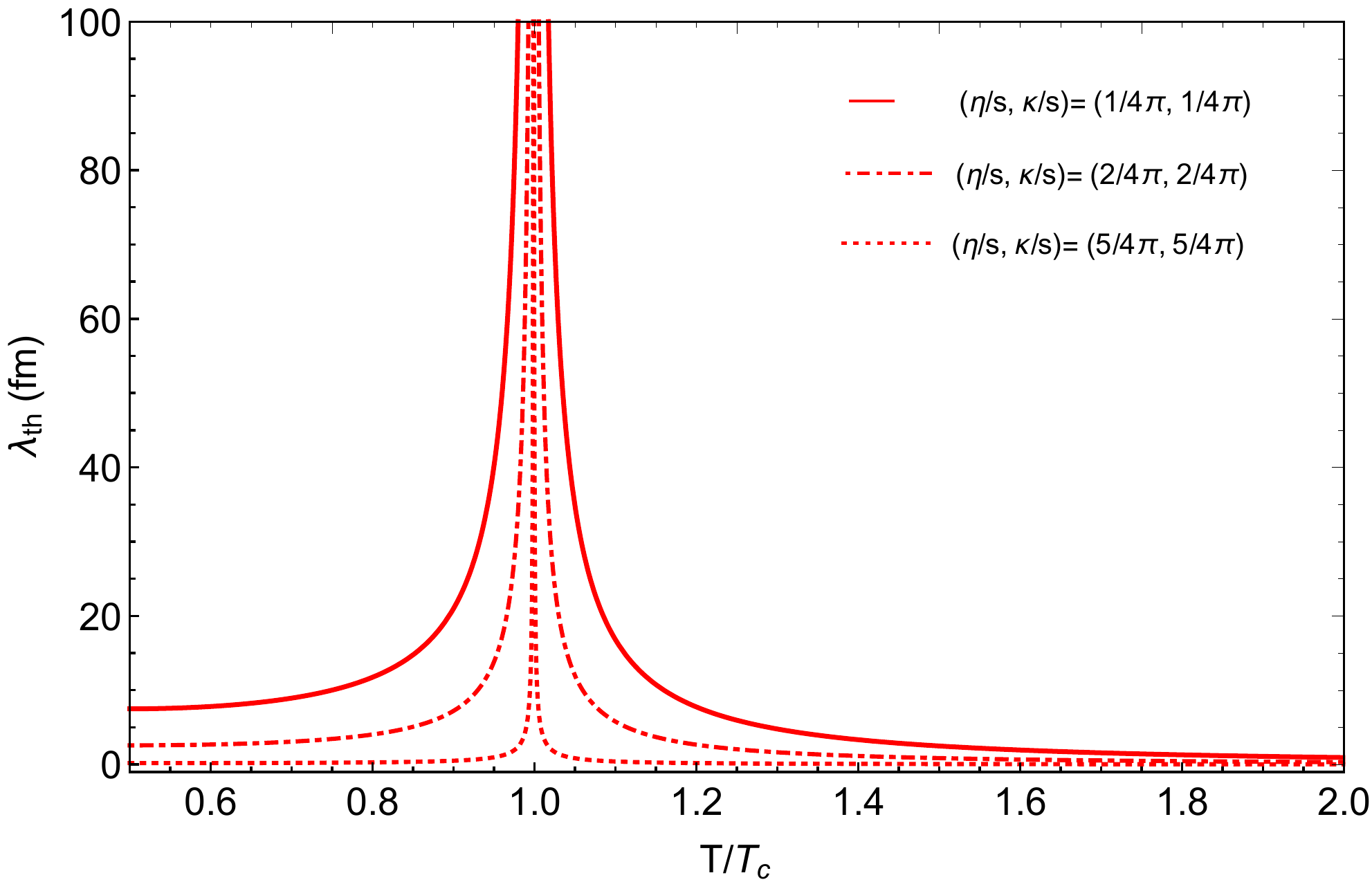}
\caption{(color online) a) Left figure is $\lambda_{th} $ (fm) vs temperature plot. Divergence is observed at $(T_{c}, \mu_{c})= (154, 367)$ MeV. b) Right plot is $\lambda_{th}$ (fm) vs T at $\mu=367 MeV$ for different sets of value transport coefficients.}
\label{fig4}
\end{figure}
Now we discuss the dissipation of the perturbation in the fluid when it hits the CEP in the QCD
phase diagram. 
The damping caused by the imaginary part of the frequency of hydrodynamic modes of perturbation 
at the critical point $(T_c, \mu_c)$ is shown in Fig.\ref{fig2}.  It is clearly seen 
from the figure that the waves with larger (smaller) values of wavenumber ($k$) damp faster (slower). 
The waves in fluid damp faster for larger values of transport coefficients 
(right panel).  
Away from the critical point the waves damp slower for high fluid temperature and density  
as evident from the results displayed in Fig.\ref{fig3} (left panel). 
The waves in a medium with higher viscosity and thermal conductivity
damp faster (Fig.~\ref{fig3}, right panel) for obvious reasons.

The variation of $\lambda_{th}$ with temperature ($T/T_c$) is shown in Fig.\ref{fig4}. 
The value of the wavelength ($\lambda_{th}=2\pi/{k_{th}}$) depends on the transport 
coefficients ($\eta, \kappa$) as well as on the various response functions appearing
through the derivatives, 
$\Big(\frac{\partial \epsilon}{\partial T}\Big)_n,\Big (\frac{\partial \epsilon}{\partial n}\Big)_T,
\Big(\frac{\partial p}{\partial T}\Big)_n, \Big(\frac{\partial p}{\partial n}\Big)_T, \Big(\frac{\partial n}{\partial \mu}\Big)_T, \Big(\frac{\partial n}{\partial T}\Big)_\mu$, 
relaxation coefficients ($\beta_{1}, \beta_2$) and the coupling constant ($\alpha_1$).  
The transport coefficients are taken as $\eta/s=\kappa/s=1/4\pi$. 
The values of $\Big(\frac{\partial \epsilon}{\partial T}\Big)_n,
\Big (\frac{\partial \epsilon}{\partial n}\Big)_T,
\Big(\frac{\partial p}{\partial T}\Big)_n, \Big(\frac{\partial p}{\partial n}\Big)_T, \Big(\frac{\partial n}{\partial \mu}\Big)_T, \Big(\frac{\partial n}{\partial T}\Big)_\mu$
are calculated in terms of different response function by
using relevant thermodynamic relations (Appendix A).  
In the left panel of Fig.\ref{fig4} the variation of $\lambda_{th}$  with $T/T_c$ is depicted.
It is observed that at CEP ($T_c= 154 \text{MeV}$, $\mu_{c}= 367 \text{MeV}$) the $\lambda_{th}$ 
diverges. As mentioned above $\lambda_{th}$ is defined as the threshold wavelength {\it i.e.} 
waves with  wavelengths, $\lambda\geq \lambda_{th}$ are allowed to propagate and
others dissipated. 
It is noted that when we consider $\mu= 347$ MeV and 387 MeV (away from critical point)
a finite value of $\lambda_{th}$ is obtained, that is wave with $\lambda>\lambda_{th}$ will
propagate in the medium without substantial dissipation in such cases. 
At the critical point, however, $\lambda_{th}$ diverges which imply that  waves with
any finite wavelength will dissipate strongly in the fluid. 
We also observe that for lower value of $T$ the value of $\lambda_{th}$ is smaller.
For $\mu= 387$ MeV  the magnitude of $\lambda_{th}$ is larger compared to $\mu = 347$ MeV. 
This indicates that for higher values of temperature and chemical potential  $\lambda_{th}$
is higher.
The fluidity defined in Eq.\eqref{eq32} is directly proportional to $\lambda_{th}$ which 
diverges at CEP, implies that fluidity also diverges at the CEP. Away from CEP, 
the fluidity decreases.  The fluidity is larger for $\mu= 387$ MeV compared to $\mu= 347$ MeV.\\\\
The viscous damping of perturbation can be understood
from the relation: $T^{\mu \lambda}_{1}(t)= T^{\mu\lambda}_{1}(0) exp({-\omega_{\Im m}t})$,
where $T^{\mu \lambda}_{1}(0)$ is the perturbation in EMT at $t=0$
and $T^{\mu\lambda}_{1}(t)$ at some later time $t$ which
is dissipated as indicated by the exponential term.
The spectrum of initial $(t=0)$ perturbations can be associated with the harmonics of the 
shape deformations and density fluctuations \cite{lacey}. 
The dispersion relation for $\omega$ provides the value, $k_{th}$, which can
be used to define a length scale, $R_v\sim 1/k_{th}$. For system of size $R$, $R_v$ can be used to define
$n_{v}= \frac{2\pi R}{R_{v}}$ which is linked to the value of the highest harmonic $n_v$ 
(eccentricity-driven) that will effectively survive damping. 
We have seen that the nature of the plot i.e $\lambda_{th}$ vs $T$ does not change much with 
the variation of the shear viscosity ($\eta/s$) but changes significantly with 
the variation of thermal conductivity ($\kappa/s$). 
Right panel of Fig.\ref{fig4} shows the variation of 
$\lambda_{th}$ with temperature for higher values of $\kappa/s$ and $\eta/s$. As the magnitude of 
$\kappa/s$ increases the gap between the divergences in the two phases gets narrower. 
It  is well-known~\cite{Kapusta} 
that the thermal conductivity diverges at critical point.  Therefore, the nature of the variation
of $\lambda_{th}$ with $T/T_c$ near the CEP will be essentially governed by the 
convergence of the thermal conductivity. 
We have found that  with increasing thermal conductivity the width of the divergence gets narrower.

\section{Summary and conclusion}
We have constructed an EoS of a fluid which contain the effects of QCD critical point and used
it to study the propagation of sound wave through the medium.
A perturbation has been imparted on the relativistic fluid and its evolution has been 
studied as it passes through the CEP within the scope of IS like causal hydrodynamics. 
We have estimated the threshold value of the wavelength, $\lambda_{th}$ 
such that any wave with wavelength below $\lambda_{th}$ is dissipated for given
values of transport coefficients and other thermodynamic quantities.
Most interestingly we have found that no waves is allowed to propagate
if the system hits the CEP {\it i.e} waves with all wavelength get dissipated at CEP
irrespective to the values of transport coefficients.
The fluidity of the system diverges at the  CEP indicating the fact that
the fluid flows without any resistance.  

It has been observed experimentally in conventional condensed matter system~\cite{WGS}
(see also~\cite{attenuationsound})
that the absorption is maximum due to diffraction of
sound from the critical region similar to the scattering of light at the critical point 
where the opalescence due to critical phenomena is strongest.
Therefore, the absorption of sound will indicate 
the presence of critical point. 
Near the critical point the correlation length ($\xi$) becomes very large, therefore,
the hydrodynamic limit, $\xi<<\lambda$ is violated. As a consequence the development
of sound wave is prevented. The forbiddance of sound wave will lead to the vanishing 
of Mach cone  (Mach angle, $\alpha=sin^{-1}(c_s/v)$, $v$ is the fluid velocity). Therefore,
the vanishing of Mach angle will indicate the presence of critical point.

Various harmonics of the azimuthal distribution of produced particles
in RHIC-E are useful quantities to characterize the matter. 
For example the triangular flow helps in understanding the initial fluctuations 
and elliptic flow can be used to comprehend  the equation of state of the system. 
The presence of critical point makes the viscous horizon  scale, $R_v\sim 1/k_{th}$ to diverge.
Since the highest order of harmonics that survives varies as, $n_v\sim 2\pi R/R_v$,
ideally the vanishing harmonics will indicate the presence of critical point.   
However, the experimentally measure quantities are superpositions
of different temperatures and densities from the formation to the freeze-out stage,
therefore, even if the system hits the critical point in the $T-\mu$ plane, 
the harmonics may not vanish, but the critical point may weaken them.

The possibility of the existence and detection of 
CEP has been studied in ~\cite{Kapusta}.  The mode-mode coupling theory has been used to estimate
the thermal conductivity at the points near and away from the QCD critical
point in the $\mu-T$ plane and shown that the thermal conductivity
diverges at the critical point.  It has also been
demonstrated that the   sharp change in thermal conductivity
at the critical point is strongly reflected in  the two particle correlation
of fluctuations in rapidity space, therefore, paving the way to
confirm the existence and location of the CEP. 

In a realistic scenario the possibility of the trajectories passing through the
critical point, {\it i.e.} the trajectories hitting the $(\mu_c,T_c)$ point
in the $(\mu,T)$ plane is remote, which limits the magnitude of the fluctuations
near the critical point. These fluctuations  will remain  out of equilibrium 
due to the expansion of the system and critical slowing down~\cite{stephanov3}. 
These issues has been considered in ~\cite{yakamatsu} while studying the 
evolution of hydrodynamic fluctuations of the system formed in RHIC-E. 
The  appearance of the Kibble-Zurek length scale and its connection with 
short range spatial correlations has been discussed. 
It has also been shown that the non-flow correlations 
get enhanced in presence of critical point
and such correlations should be measured as a function of $n/s$ for detecting
the CEP~\cite{yakamatsu}.

In a realistic scenario the matter formed in RHIC-E 
evolves in space and time - from the initial QGP phase to the final hadronic 
freeze-out state through a phase transition in the intermediate stage. 
The space time evolution of the locally equilibrated system is described
by relativistic viscous hydrodynamics. The experimentally  detected signals
is the superposition of the yields for all the possible values 
of temperatures and densities  of the system ranging from the initial to freeze-out states.
The detection of CEP will require the disentanglement of contributions from the  
neighbourhood ($\mu_c,T_c$) from all other possible values of $\mu$ and $T$ which the system 
confronts during its evolution history from the initial to the freeze-out stages. 
In the present work the expansion dynamics has not been taken into consideration, therefore,
the results obtained here can not be contrasted with experiments. The effects of the CEP with
(3+1) dimensional expansion within the scope of second order viscous hydrodynamics will be 
published in future~\cite{sksingh}.

Rigorously speaking hydrodynamics is applicable in the region where 
$k<<\xi^{-1}$ is satisfied where $k$ is the wave vector of the sound
mode and $\xi$ is correlation length. At the CEP this fundamental assumptions
on the application of hydrodynamics becomes invalid as $\xi$ diverges
as the system approaches CEP with $T\rightarrow T_c$ and $\mu\rightarrow \mu_c$.
However, for a given $k$ there will certainly be a  domain in the neighbourhood 
of $(\mu_c,T_c)$ where the predictions of hydrodynamics can be useful. 
Existence of such region in condensed matter  system has been discussed in~\cite{Stanley}.

\section*{Acknowledgement} We are grateful to Golam Sarwar for useful discussions. M.H. would like 
to thank VECC for support. M.R. is supported by Department of Atomic Energy (DAE), 
Govt. of India.  

\section*{Appendix A}
The expressions for $\omega_{\Re e}(k)$ and $\omega_{\Im m}(k)$ contain derivatives of several thermodynamics quantities. In this appendix
we recast these derivatives
in terms of response functions like: isothermal and adiabatic compressibilities  ($\kappa_T$ and $\kappa_s$), volume expansivity $\alpha_{p}$ 
specific heats ($c_p$ and $c_v$), baryon number susceptibility ($\chi_{B}$), velocity of sound ($c_s$), etc. The 
baryon number density ($n$) and the entropy density ($s$)  are given by
\beqa
n=\Big(\frac{\partial p}{\partial \mu}\Big)_{T};\,\,\,\,\, s=\Big(\frac{\partial p}{\partial T}\Big)_{\mu}
\eeqa
Baryon number susceptibility ($\chi_{B}$), isothermal compressibility ($\kappa_{T}$), adiabatic compressibility ($\kappa_{s}$) and volume expansivity ($\alpha_{p}$) are given by,
\beqa
\chi_{B}=\Big(\frac{\partial n}{\partial \mu}\Big)_{T}; \kappa_{T}=\frac{1}{n}\Big(\frac{\partial n}{\partial p}\Big)_{T}; \kappa_{s}=\frac{1}{n}\Big(\frac{\partial n}{\partial p}\Big)_{s}; \alpha_{p}=\frac{1}{V}\Big(\frac{\partial V}{\partial T}\Big)_{p}=-\frac{1}{n}\Big(\frac{\partial n}{\partial T}\Big)_{p}
\eeqa
Specific heats are given by
\beqa
c_{p}=T\Big(\frac{\partial s}{\partial T}\Big)_{p}; c_{V}=T\Big(\frac{\partial s}{\partial T}\Big)_{V}=T\Big(\frac{\partial s}{\partial T}\Big)_{n}=\Big(\frac{\partial \epsilon}{\partial T}\Big)_{V}=\Big(\frac{\partial \epsilon}{\partial T}\Big)_{n}
\eeqa
We  have to express six quantities such as: $(\frac{\partial p}{\partial T})_{n}, 
(\frac{\partial p}{\partial n})_{T}, (\frac{\partial \epsilon}{\partial T})_{n}, (\frac{\partial n}{\partial T})_{\mu}, (\frac{\partial n}{\partial \mu})_{T}$ and
$\Big(\frac{\partial \epsilon}{\partial n}\Big)_{T}$. \\
i) To evaluate: $(\frac{\partial p}{\partial T})_{n}$ we start with
\beqa
\Big(\frac{\partial p}{\partial T}\Big)_{n}&=&\frac{\pd (p, n)}{\pd (T,n)}
=\frac{\pd (p, n)}{\pd (T, p)}\frac{\pd (T, p)}{\pd (s, p)}\frac{\pd (s, p)}{\pd (s, \epsilon)}\frac{\pd (s, \epsilon)}{\pd (s, n)}\frac{\pd (s, n)}{\pd (T, n)} \nn\\
&=& \Big[-\Big(\frac{\partial n}{\partial T}\Big)_{p}\Big]\Big(\frac{\partial T}{\partial s}\Big)_{p}\Big(\frac{\partial p}{\partial \epsilon}\Big)_{s}\Big(\frac{\partial \epsilon}{\partial n}\Big)_{s}\Big(\frac{\partial s}{\partial T}\Big)_{n} \nn\\
&=& n\alpha_{p}\frac{T}{c_{p}}c^{2}_{s}\Big(\frac{\partial \epsilon}{\partial n}\Big)_{s}\frac{c_{V}}{T}
=nc^{2}_{s}\alpha_{p} \frac{c_{V}}{c_{p}}\Big(\frac{\partial \epsilon}{\partial n}\Big)_{s}
\eeqa
By using the relation,
\beqa
d\epsilon=Tds+\mu dn \,\,\,\,\, \text{and } \mu=\Big(\frac{\partial \epsilon}{\partial n}\Big)_{s}
\eeqa
we can write  \vspace{-.85cm}
\beqa
{\Big(\frac{\partial p}{\partial T}\Big)_{n}=\mu nc^{2}_{s}\alpha_{p} \frac{c_{V}}{c_{p}}}
\eeqa 

ii)  Now consider $\Big(\frac{\partial p}{\partial n}\Big)_{T}$: \vspace{-1cm}
\beqa
{\Big(\frac{\partial p}{\partial n}\Big)_{T}= \frac{1}{n\kappa_{T}}}
\eeqa
\beqa
\Big(\frac{\partial p}{\partial n}\Big)_{T}&=&\frac{\pd (p, T)}{\pd (n, T)}
=\frac{\pd (p, T)}{\pd (p, s)}\frac{\pd (p, s)}{\pd (\epsilon, s)}\frac{\pd (\epsilon, s)}{\pd (n,s)}\frac{\pd (n, s)}{\pd (n, T)}\nn\\
&=&\Big(\frac{\partial T}{\partial s}\Big)_{p}\Big(\frac{\partial p}{\partial \epsilon}\Big)_{s}\Big(\frac{\partial \epsilon}{\partial n}\Big)_{s}\Big(\frac{\partial s}{\partial T}\Big)_{n}\nn\\
&=& \frac{T}{c_{p}}c^{2}_{s}\Big(\frac{\partial \epsilon}{\partial n}\Big)_{s}\frac{c_{V}}{T}
\\&=&\mu c^{2}_{s} \frac{c_{V}}{c_{p}}
\eeqa
 
iii) The factor, $\Big(\frac{\partial \epsilon}{\partial T}\Big)_{n}$ can be estimates as follows:
\beqa
\Big(\frac{\partial \epsilon}{\partial T}\Big)_{n}=c_n
\eeqa
For fixed net baryon number,  $c_n$ can be written as $c_{n}=c_{V}$.
Therefore, 
\beqa
{\Big(\frac{\partial \epsilon}{\partial T}\Big)_{n}=c_{V}}
\eeqa
iv) $\Big(\frac{\partial \epsilon}{\partial n}\Big)_{T}$ can be estimates as:
\beqa
\Big(\frac{\partial \epsilon}{\partial n}\Big)_{T}&=&\frac{\pd (\epsilon, T)}{\pd (n,T)}
=\frac{\pd (\epsilon, T)}{\pd (\epsilon,s)}\frac{\pd (\epsilon, s)}{\pd (n,s)}\frac{\pd (n, s)}{\pd (n,T)}\nn\\
&=& \Big(\frac{\partial T}{\partial s}\Big)_{\epsilon}\Big(\frac{\partial \epsilon}{\partial s}\Big)_{s}\Big(\frac{\partial s}{\partial T}\Big)_{n} \nn\\
&=& \Big(\frac{\partial T}{\partial s}\Big)_{\epsilon}\Big[\Big(\frac{\partial \epsilon}{\partial p}\Big)_{s}\Big(\frac{\partial p}{\partial n}\Big)_{s}\Big] \frac{c_{V}}{T} \nn\\
&=& \Big(\frac{\partial T}{\partial s}\Big)_{\epsilon}\Big[\frac{1}{c^{2}_{s}}\frac{1}{n\kappa_{s}}\Big]\frac{c_{V}}{T}
= \frac{T}{c_{\epsilon}}\Big[\frac{1}{c^{2}_{s}}\frac{1}{n\kappa_{s}}\Big]\frac{c_{V}}{T}
\\&=&\frac{c_{V}}{c_{\epsilon}}\Big[\frac{1}{c^{2}_{s}}\frac{1}{n\kappa_{s}}\Big]
\eeqa
v) The quantity $\Big(\frac{\partial n}{\partial \mu}\Big)_{T}$ can be estimates as
\beqa
\Big(\frac{\partial n}{\partial \mu}\Big)_{T} =\chi_{B}
\eeqa
vi) For $\Big(\frac{\partial n}{\partial T}\Big)_{\mu}$, we have,
\beqa
\Big(\frac{\partial n}{\partial T}\Big)_{\mu}=-\Big(\frac{\partial n}{\partial \mu}\Big)_{T}\Big(\frac{\partial \mu}{\partial T}\Big)_{n}=-\chi_{B}\Big(\frac{\partial \mu}{\partial T}\Big)_{n}
\eeqa
We know that,
\beqa
sdT&=&dp-nd\mu \to dp=sdT+nd\mu \\
Tds&=&d\epsilon-\mu dn \to d\epsilon =Tds+\mu dn\\
\Big(\frac{\partial p}{\partial \epsilon}\Big)&=&\frac{sdT+nd\mu }{Tds+\mu dn}\\
&=& \frac{s+n\Big(\frac{\partial \mu}{\partial T}\Big)}{T\Big(\frac{\partial s}{\partial T}\Big)+\mu\Big(\frac{\partial n}{\partial T}\Big)}\\
n\Big(\frac{\partial \mu}{\partial T}\Big)&=&\Big(\frac{\partial p}{\partial \epsilon}\Big)\Big[T\Big(\frac{\partial s}{\partial T}+\mu \Big(\frac{\partial n}{\partial T}\Big)\Big]-s
\eeqa
Thus
\beqa
\Big(\frac{\partial \mu}{\partial T}\Big)_{n}=\frac{c^{2}_{n}C_{V}-s}{n} \hspace*{1cm} \text{where}, c^{2}_{n}=\Big(\frac{\partial p}{\partial \epsilon}\Big)_{n}
\eeqa
Finally we get,
\beqa
\Big(\frac{\partial n}{\partial T}\Big)_{\mu}=\frac{\chi_{B}}{n}(s-c^{2}_{n}C_{V})
\eeqa


\end{document}